\documentclass[aps,prb,twocolumn,superscriptaddress]{revtex4-1}
\usepackage{amsmath,amsthm,amssymb}
\usepackage[dvipdfmx]{graphicx}
\usepackage{amsmath,bm}
\usepackage{color,ulem}
\usepackage{ifthen}
\usepackage{mdwlist}
\newcount \commentout
\newcommand{\1}{\mbox{1}\hspace{-0.25em}\mbox{l}}

\newlength{\figwidth}
\newlength{\figlarge}
\begin{document}
\title{
Non-Hermitian perspective of the band structure in heavy-fermion systems
}
\author{Tsuneya Yoshida}
\affiliation{Department of Physics, Kyoto University, Kyoto 606-8502, Japan}
\affiliation{Department of Physics, University of Tsukuba, Ibaraki 305-8571, Japan}
\author{Robert Peters}
\affiliation{Department of Physics, Kyoto University, Kyoto 606-8502, Japan}
\author{Norio Kawakami}
\affiliation{Department of Physics, Kyoto University, Kyoto 606-8502, Japan}
\date{\today}
\begin{abstract}
We analyze a two-dimensional Kondo lattice model with special emphasis on non-Hermitian properties of the single-particle spectrum, following a recent proposal by Kozii and Fu.
Our analysis based on the dynamical mean-field theory elucidates that the single-particle spectral weight shows the exceptional points (EPs). Correspondingly, the spectral weight exhibits the band touching, resulting in a structure similar to the ``Fermi arc". Furthermore, we find an intriguing phenomenon arising from the periodicity of the lattice. The EPs generated by two distinct Dirac points merge and change into a hybrid point which vanishes as the exchange interaction is increased. Accordingly, the paramagnetic phase in the low temperature region shows a significant difference from non-interacting fermions: the imaginary part of the self-energy yields the ``Fermi loop" without any defective points.
\end{abstract}
\pacs{
***
} 
\maketitle

\section{introduction}

The importance of the topological perspective in condensed matter systems is rapidly growing. 
In particular, topological systems are extended to superconductors~\cite{Kitaev_chain_01,Majorana_Mourik,Majorana_Rokhinson2012,Majorana_Anindya2012} and semi-metals~\cite{XWan_PRB11_Weyl,AABurkov_PRL11_Weyl,HWeng_PRX15_Weyl,SYXu_Science15_Weyl,BQLv_PRX15_Weyl} where topologically nontrivial properties induce robust degenerate states, providing platforms for novel excitations in solids, such as Weyl fermions and Majorana fermions. 
In particular, the realization of Majorana fermions attracts much interest in terms of application to quantum computation. 
Great progress in topological systems has also brought impact on strongly correlated systems 
where various intriguing phenomena have been reported because of topology and correlations~\cite{MHohenadler_PRL11_corr_topo,Yamaji11,SLYu_PRL11_corr_topo,TBI_Mott_Yoshida,TBI_Mott_Tada,WWu_PRB12_corr_topo,Yoshida_PRB13_AFTI}; the topological Mott insulators emerge whose topology is reflected only in collective gapless edge modes~\cite{TMI_LBalents09,TMI_YoshidaPRL14,TMI_YoshidaPRB16,HQWuPRB16_TMI}; 
electron correlations reduce the $\mathbb{Z}$ classifications for free fermions~\cite{Z_to_Zn_Fidkowski_10,Fidkowski_1Dclassificatin_11,Turner11,Ryu_Z_to_Z8_2013,YaoRyu_Z_to_Z8_2013,Qi_Z_to_Z8_2013,Fidkowski_Z162013,Hsieh_CS_CPT_2014,Isobe_Fu2015,Yoshida2015,Wang_Potter_Senthil2014,You_Cenke2014,Wang_Senthil2014,Morimoto_2015,Yoshida_PRL17_ZtoZ8,Yoshida_arXiv18_ZtoZ4}. 
The destruction of the degenerate states by the electron correlations is the origin of these phenomena.

Intriguingly, a very recent study by Kozii and Fu has proposed a new mechanism of robust degenerate states described by a non-Hermitian matrix for electron systems in equilibrium~\cite{VKozii2017_non-Hermi}. 
They have found robust degenerate states induced by electron correlations, and have clarified their topological properties.
The single-particle spectrum of the multi-band systems is described by a non-Hermitian matrix due to the imaginary part of the self-energy, describing quasi-particle lifetime. 
Remarkably, in non-Hermitian systems, the diagonalizability of the matrix can be violated at a certain point in the Brillouin zone (BZ) which is denoted as EP.
At this point, the eigenvalues of the non-Hermitian matrix are degenerate, which is robust and does not require any symmetry~\cite{TELeePRL16_Half_quantized,YXuPRL17_exceptional_ring,HShen2017_non-Hermi}.
The EPs in the BZ are connected with lines where the single-particle spectrum shows band touching, resulting in the ``Fermi arc".
The proposal in Ref.~\onlinecite{VKozii2017_non-Hermi} bridges two distinct issues of condensed matter, electron systems in equilibrium and systems described by a non-Hermitian matrix where various intriguing properties have been reported so far~\cite{NHatano_PRL96,CMBender_PRL98,BZhen_Nat15,ZPGong_PRL17,YAshida_NatCom17,KKawabata_arXiv18,ZPGong_arXiv18}.
After this proposal, systems with disorder have been extensively studied as platforms of the new robust degenerate states induced by the self-energy~\cite{HShen2018quantum_osci,Papaj2018bulk_oci}. These robust degenerate states also attract interest because of its potential to solve the puzzle of quantum oscillations in $\mathrm{SmB_6}$ and $\mathrm{YbB_{12}}$~\cite{BTan_Science15_OscillationSmB6,ZXiang_underreview_OscillationYbB12,HLiu_IOP18_OscillationYbB12,HShen2018quantum_osci}. 

In spite of the extensive studies, strongly correlated electron systems where the self-energy plays an important role have not been sufficiently explored yet from the non-Hermitian perspective.

Under this background, we here study strongly correlated electron systems as an arena for the degenerate states arising from non-Hermitian properties, which provides a new direction in the study of correlated topological systems.
In particular, we analyze a heavy-fermion system in two dimensions by employing the dynamical mean-field theory combined with the numerical renormalization group method (DMFT+NRG). 
Our analysis elucidates that the imaginary part of the self-energy splits the Dirac point into two EPs with vorticity $\nu=\pm 1/2$. Furthermore, we find an intriguing phenomenon arising from the periodicity of the lattice. The EPs generated by two distinct Dirac points merge and change to a hybrid point which vanishes when the Kondo temperature becomes large. Remarkably, the paramagnetic phase in the low temperature region shows a significant difference from non-interacting fermions: the imaginary part of the self-energy yields a ``Fermi loop" where the generalized charge gap becomes pure imaginary.

The rest of this paper is organized as follows. In Sec.~\ref{sec: model and method}, we describe our setup and give a brief explanation of our approach.
In Sec.~\ref{sec: results}, we observe the emergence of EPs and the fusion of them induced by the Kondo effect.
The last section is devoted to a brief summary.

\section{model and method}
\label{sec: model and method}

We consider the following Kondo lattice model on the square lattice
\begin{eqnarray}
\label{eq: Hami_KLM}
H &=& \sum_{\langle i j\rangle s \alpha \beta } t_{i\alpha,j\beta}c^\dagger_{i \alpha s} c_{j \beta s} +\sum_{i} J\bm{s}_i \cdot \bm{S}_i,
\end{eqnarray}
where $c^\dagger_{i\alpha s}$ creates an electron with spin $s=\uparrow,\downarrow$ at orbital $\alpha=a,b$ of site $i$. $\bm{s}_i:=c^\dagger_{ibs} \bm{\sigma}_{ss'} c^\dagger_{ibs'}/2$
$\sigma$'s are the Pauli matrices acting on the spin space. $\bm{S}$ is the spin 1/2 operator of the local spin.
The first term describes the non-interacting part which is written in the momentum space as follows:
\begin{eqnarray}
\hat{h}(\bm{k}) &=&
d_z(\bm{k})\tau_z+d_x(\bm{k})\tau_x,
\end{eqnarray}
with $d_z(\bm{k})=-\epsilon_0-2t(\cos k_x +\cos k_y)$ and $d_x(\bm{k})=2t_{sp}\sin k_y$.
The Pauli matrices $\tau$'s act on the orbital space.
In the non-interacting case, this system hosts Dirac cones at points in the BZ where $d_x(\bm{k})$ and $d_z(\bm{k})$ become zero.
These Dirac points are protected by the chiral symmetry with $\tau_y \hat{h}(\bm{k}) \tau_y =- \hat{h}(\bm{k})$. 
We note that only electrons at orbital $b$ interact with the localized spins.
In general, the coupling strength depends on the details of the itinerant orbital and thus can be different for each orbital. 
In this study, we take the extreme case where only electrons in one of the orbitals couple with the localized spins.

In order to analyze correlation effects, we employ the dynamical mean-field theory (DMFT) which treats local correlations exactly~\cite{WMetznerPRL89_DMFT,MHartmannZP89_DMFT,AGeorgesRMP96_DMFT}.
In the DMFT framework, the lattice model is mapped onto the following effective impurity model
\begin{subequations}
\begin{eqnarray}
 \mathcal{Z}_{\mathrm{imp}} &=& \int \!\!\mathcal{D} \bar{c}_{b\sigma}(\tau) \mathcal{D} c_{b\sigma}(\tau) \mathrm{Tr}_{\bm{S}}\exp(- \mathcal{S}_\mathrm{imp} ),  \\
 \mathcal{S}_{\mathrm{imp}} &=& \int \!\!d\tau d\tau'  \left[ \sum_{\sigma}\bar{c}_{b\sigma}(\tau) \mathcal{G}_\sigma(\tau-\tau') c_{b}(\tau') \right.\nonumber\\
 &&\quad\quad \left.-J \bm{s}(\tau)\cdot \bm{S}\delta(\tau-\tau') \right],
\end{eqnarray}
\end{subequations}
where $\bar{c} _{b\sigma}$ is a Grassmannian variable corresponding to the creation operator at the orbital $b$ of site 0.
$\mathrm{Tr}_{\bm{S}}$ denotes the trace for states of the localized spin.
Here, we note that the electrons of the orbital $a$ are integrated out in the above effective model.
$\mathcal{G}_\sigma(\tau-\tau')$ denotes the Green's function of the effective bath which can be obtained by solving the following self-consistent equation
\begin{eqnarray}
\label{eq: self-eq_para}
\mathcal{G}^{-1}_\sigma(\omega) &=& [\sum_{\bm{k}} \{ (\omega +i\delta)\1 -h(\bm{k})-\Sigma^R_\sigma(\omega) \}^{-1} ]_{bb} \nonumber \\
&&\quad +\Sigma^R_{b\sigma}(\omega), 
\end{eqnarray}
where 
$
\Sigma^R_\sigma(\omega):=\mathrm{diag}\left(
\begin{array}{cc}
0 & \Sigma^R_{b\sigma}(\omega)
\end{array}
\right)
$, and $\Sigma^R_{b\sigma}(\omega)$ denotes the self-energy of the retarded Green's function describing electrons in orbital $b$.

In order to solve the self-consistent equation~(\ref{eq: self-eq_para}), we employ the numerical renormalization group (NRG) method which provides accurate results even around zero temperature~\cite{KWilsonRMP75_NRG,RPetersPRB06_NRG,RBullaRMP08_NRG}.
In this study, we also analyze the magnetic order with the DMFT by dividing the lattice into two sublattices.

\section{Results}
\label{sec: results}

\subsection{Defective points via spectral functions $A(\omega,\bm{k})$ and their characterization}
Here, we discuss the condition where the momentum-resolved spectral function shows physics arising from the breakdown of diagonalizability, a characteristic behavior of non-Hermitian systems.
We assume that the system is paramagnetic, i.e., $\Sigma^R_b(\omega):= \Sigma^R_{b\uparrow}(\omega)= \Sigma^R_{b\downarrow}(\omega)$.
In this case, the Green's function is written as
%
%
\begin{subequations}
\label{eq: Green_Heff}
\begin{eqnarray}
G(\omega+i\delta,\bm{k})^{-1} &=& (\omega+i\delta)\tau_0 -H_{eff}(\omega,\bm{k}),
\end{eqnarray}
with the effective Hamiltonian 
\begin{eqnarray}
H_{eff}(\omega,\bm{k}) &=& h(\bm{k})+\frac{\Sigma^R_{b}(\omega)}{2} (\tau_0-\tau_z).
\end{eqnarray}
\end{subequations}
$H_{eff}(\omega,\bm{k})$ is a non-Hermitian matrix because of the imaginary part of the self-energy.
If $H_{eff}(\omega,\bm{k})$ is not diagonalizable, the effective Hamiltonian is defective.

When $H_{eff}(\omega,\bm{k})$ is not defective, we obtain the following representation of the spectral function $A(\omega,\bm{k})$ by diagonalizing the effective Hamiltonian:
\begin{subequations}
\label{eq: Ak and Epm}
\begin{eqnarray}
A(\omega,\bm{k}) &=& -\frac{1}{\pi}\mathrm{Im} \sum_{s=\pm1} [\omega+i\delta -E_{s}(\omega,\bm{k})]^{-1},
\end{eqnarray}
with the frequency dependent eigenvalues 
\begin{eqnarray}
E_{\pm}(\omega,\bm{k})&=& \frac{\Sigma^R_{b\sigma}(\omega)}{2} \pm \sqrt{ ( d_z(\bm{k})-\frac{\Sigma^R_{b\sigma}(\omega)}{2})^2 + d^2_x(\bm{k}) }. \nonumber \\
\end{eqnarray}
\end{subequations}                 

When the effective Hamiltonian is defective, its eigenvalues are degenerate. 
This robust degeneracy can be observed via spectral weight at 
$(\omega_0,\bm{k}_0)$ specified by the following conditions 
\begin{subequations}
\label{eq: EP_cond}
\begin{eqnarray}
&&\omega_0 -\mathrm{Re}[\frac{\Sigma^R_{b\sigma}(\omega_0)}{2}]=0, \\
&& d_z(\bm{k}_0)-\mathrm{Re}[\frac{\Sigma^R_{b\sigma}(\omega_0)}{2}]=0,\\
&& -(\mathrm{Im}\Sigma^R_{b\sigma}(\omega_0))^2 +4d^2_x(\bm{k}_0) =0.
\end{eqnarray}
\end{subequations}
Eq.~(\ref{eq: EP_cond}a) specifies the energy $\omega_0$ where the robust degenerate states are observed.
The other two equations specify points in the BZ where the effective Hamiltonian~(\ref{eq: Green_Heff}b) becomes defective (see Appendix~\ref{sec: app_diag}).

We note that a topological aspect of the defective points can be characterized by the vorticity defined as~\cite{HShen2017_non-Hermi}
\begin{eqnarray}
\label{eq: vorticity}
\nu &=&  \oint \frac{d\bm{k}}{2\pi }\cdot \nabla_{\bm{k}}\mathrm{arg}[E_{+}(\bm{k})-E_{-}(\bm{k})],
\end{eqnarray}
where the integral is along a closed path in the BZ.
If the defective point is characterized by a half-integer vorticity, it is denoted as an exceptional point (EP).


\subsection{DMFT results}
In the following, we discuss the DMFT results obtained for $(t,t_{sp},\epsilon)=(1,0.667,0.667)$ where the linear dispersion relation holds for a relatively wide range of energy.
We note that similar behaviors can also be observed for other sets of parameters.

\subsubsection{
Emergence of EPs and disappearance of them in the high temperature region
}
Let us first summarize a few basic properties of Kondo lattice systems for a given temperature.
When the antiferromagnetic interaction $J$ is small, the itinerant electrons are effectively decoupled from the localized spins due to temperature effects. In this case, the electrons are essentially free fermions.
With increasing interaction $J$ one can observe that the singlet correlation of electrons and localized spins is enhanced, corresponding to the increase of the Kondo temperature.
Accordingly, the band structure of the electrons is renormalized and the imaginary part of the self-energy is enhanced.
%
\begin{figure}[!h]
\begin{minipage}{1.0\hsize}
\begin{center}
\includegraphics[width=\hsize,clip]{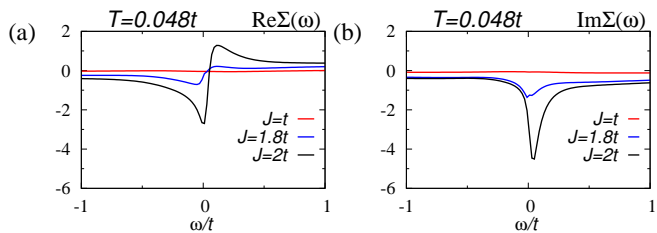}
\end{center}
\end{minipage}
\caption{
(Color Online).
The self-energy of the single-particle Green's function for several values of exchange interaction $J$. (a): the real part. (b) the imaginary part. 
As the interaction $J$ is increased, the Kondo effect is enhanced, which results in strong renormalization observed in panel (a) and a peak in the imaginary part around $\omega\sim0$ as observed in panel (b).
}
\label{fig: Sigma}
\end{figure}

The renormalization of the band structure and the increase of the imaginary part of the self-energy are observed around $J=1.8t$ for $T=0.048t$.
As seen in Fig.~\ref{fig: Sigma}, the real-part of the self-energy shows an abrupt change around $\omega\sim0$. Correspondingly, the imaginary part of the self-energy shows a dip. These structures of the self-energy induce the characteristic behaviors of the momentum resolved spectral weight $A(\omega,\bm{k})$.
%

%
\begin{figure}[!h]
\begin{minipage}{0.475\hsize}
\begin{center}
\includegraphics[width=\hsize,clip]{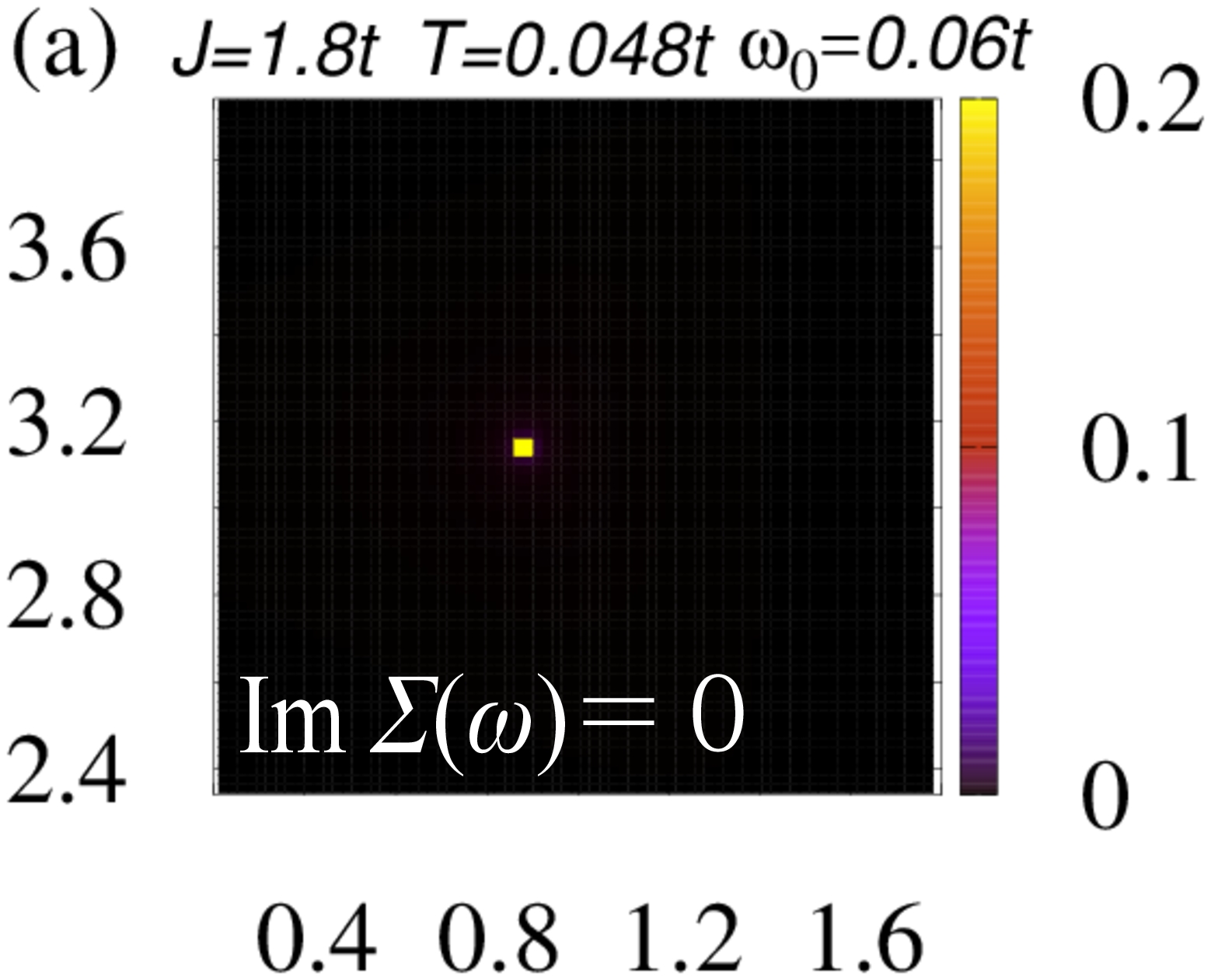}
\end{center}
\end{minipage}
\begin{minipage}{0.475\hsize}
\begin{center}
\includegraphics[width=\hsize,clip]{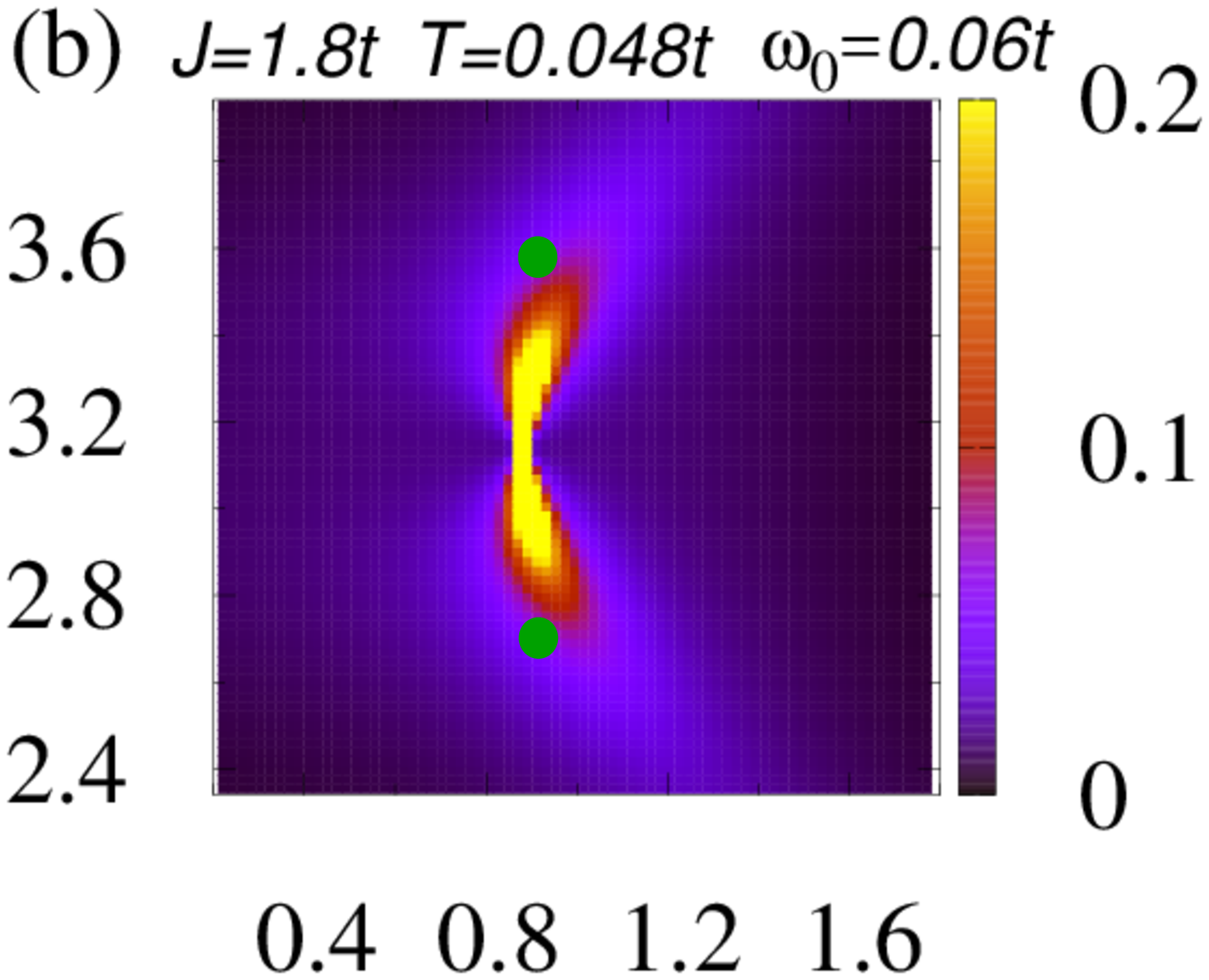}
\end{center}
\end{minipage}
\caption{(Color Online).
Momentum resolved spectral function for a given frequency $\omega_0$. 
In panel (a), we plot the spectral function by setting the imaginary part of the self-energy to zero.
We note that the spectral function is symmetric under the transformations, $(k_x,k_y)\to(-k_x, k_y)$ and $(k_x,k_y)\to(k_x, -k_y)$. 
Because of the imaginary part of the self-energy, the effective Hamiltonian~(\ref{eq: Green_Heff}b) becomes defective at $(k_{0x},k_{0y})=(0.93,2.71)$ which is denoted as a green dot in panel (b). 
}
\label{fig: Ak_EP}
\end{figure}
\begin{figure}[!h]
\begin{minipage}{0.475\hsize}
\begin{center}
\includegraphics[width=\hsize,clip]{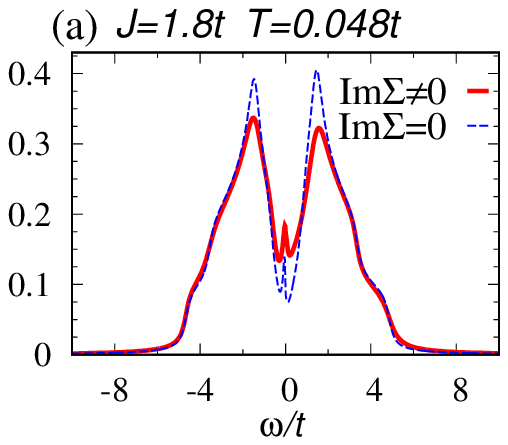}
\end{center}
\end{minipage}
\begin{minipage}{0.475\hsize}
\begin{center}
\includegraphics[width=\hsize,clip]{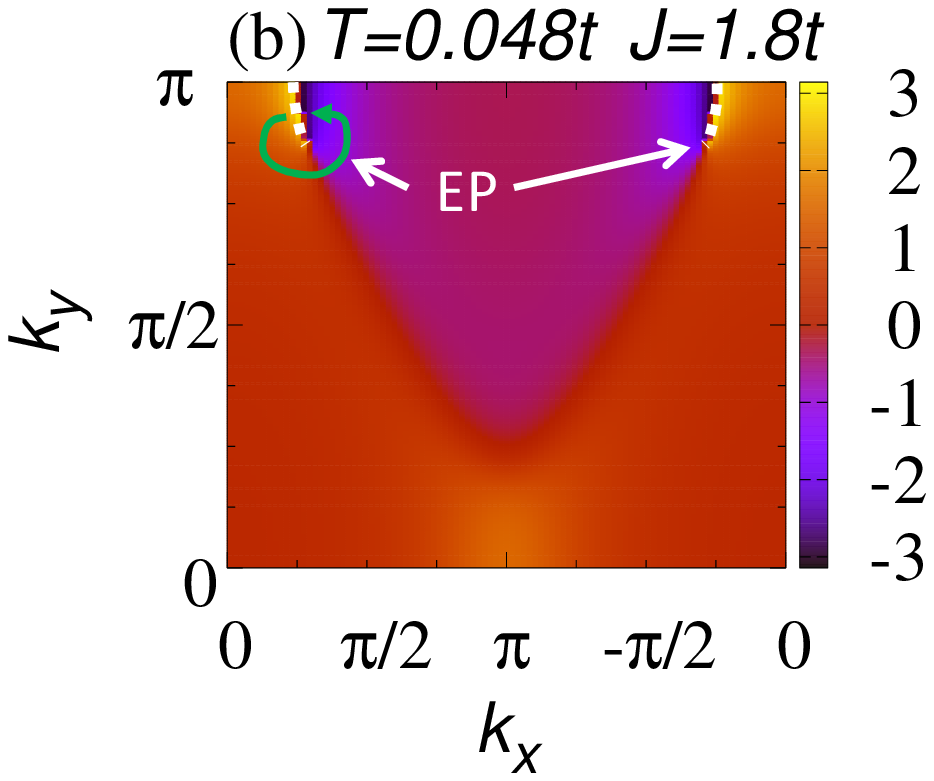}
\end{center}
\end{minipage}
\caption{
(Color Online).
(a): Local density of states for $J=1.8t$. 
The dashed lines represent the data obtained by setting $\mathrm{Im}\Sigma^R_{b}(\omega)=0$. 
The local density of states is enhanced by the imaginary part of the self-energy which induces the ``Fermi arc" observed for the momentum resolved spectral functions.
(b): Color map of $\Delta^2_c$. The dashed white lines denote the branch cut.
The white arrows point at EPs. The green line with arrows denotes the path of the integral of Eq.~(\ref{eq: vorticity}).
}
\label{fig:LDOS_Color_J1.8}
\end{figure}

To see this, we first analyze the renormalization effects of the band structure by setting the imaginary part to zero.
We note that the spectral function is symmetric under transformations $(k_x,k_y)\to(-k_x,k_y)$ and $(k_x,k_y)\to(k_x,-k_y)$.
In Fig.~\ref{fig: Ak_EP}(a), the spectral function shows a peak at $k_y=\pi$ which corresponds to the Dirac cone.
Next, we analyze the effect of the imaginary part of the self-energy on the spectral function. 
Due to the imaginary part, the effective Hamiltonian~(\ref{eq: Green_Heff}) becomes defective at $\bm{k}\sim(k_{0x},k_{0y}):=(0.93,2.71)$ and $(k_{0x},-k_{0y})$ denoted with green dots in Fig.~\ref{fig: Ak_EP}(b).
Taking into account the symmetry of the spectrum we observe four defective points which are connected with low energy excitations forming a structure similar to the Fermi arc [see Fig.~\ref{fig: Ak_EP}(b)]. The origin of the ``Fermi arc" can be understood by examining Eq.~(\ref{eq: Ak and Epm}). 
This equation indicates that only the real part of $E_{\pm}$ governs the position of the peak. Thus, we can understand that the band touching of the spectral function, leading to the ``Fermi arc", is observed along the line where the generalized charge gap 
\begin{eqnarray}
 \Delta_c(\bm{k})&=&E_{+}(\bm{k})-E_{-}(\bm{k}),
\end{eqnarray}
becomes pure imaginary.
We note that the emergence of the ``Fermi arc" enhances the weight of the local density of states around $\omega\sim0$ [see Fig.~\ref{fig:LDOS_Color_J1.8}(a)].
%

Next, for the characterization of the defective point, we compute the vorticty $\nu$. 
In Fig.~\ref{fig:LDOS_Color_J1.8}(b), the square of the charge gap $\Delta^2_c$ is plotted [see also Eq.~(\ref{eq: vorticity})].
The integral is taken along the green line in the figure. We can observe that the path of the integral crosses the branch cut once, which results in the vorticity $\nu=-1/2$. Thus, this defective point is an EP with $\nu=-1/2$.
In a similar way, we find three other EPs; an EP with $\nu=-1/2$ at $(-k_{0x},-k_{0y})$ and EPs with $\nu=1/2$ at $(k_{0x},-k_{0y})$ and $(-k_{0x},k_{0y})$.

\begin{figure}[!h]
\begin{minipage}{0.475\hsize}
\begin{center}
\includegraphics[width=\hsize,clip]{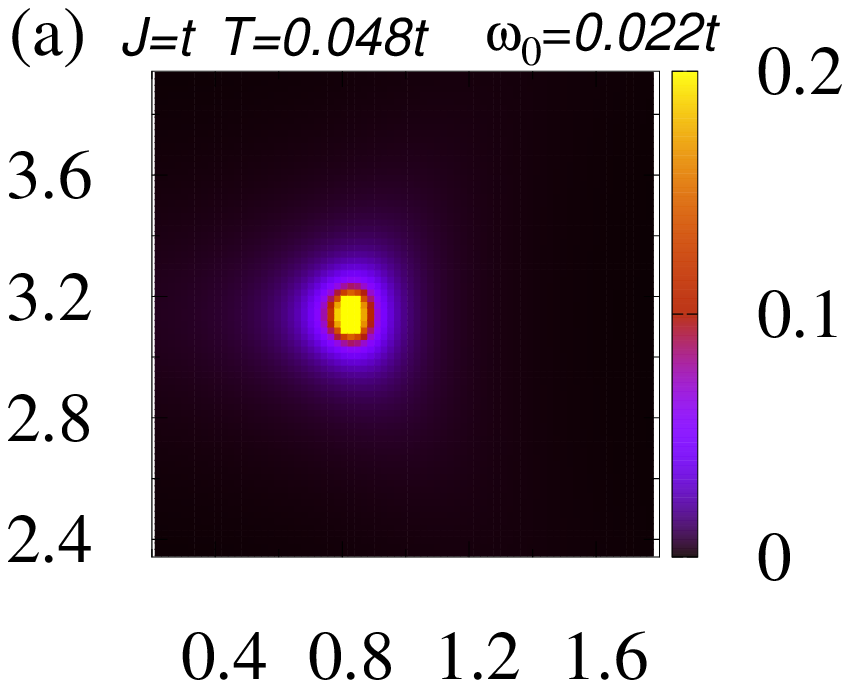}
\end{center}
\end{minipage}
\begin{minipage}{0.475\hsize}
\begin{center}
\includegraphics[width=\hsize,clip]{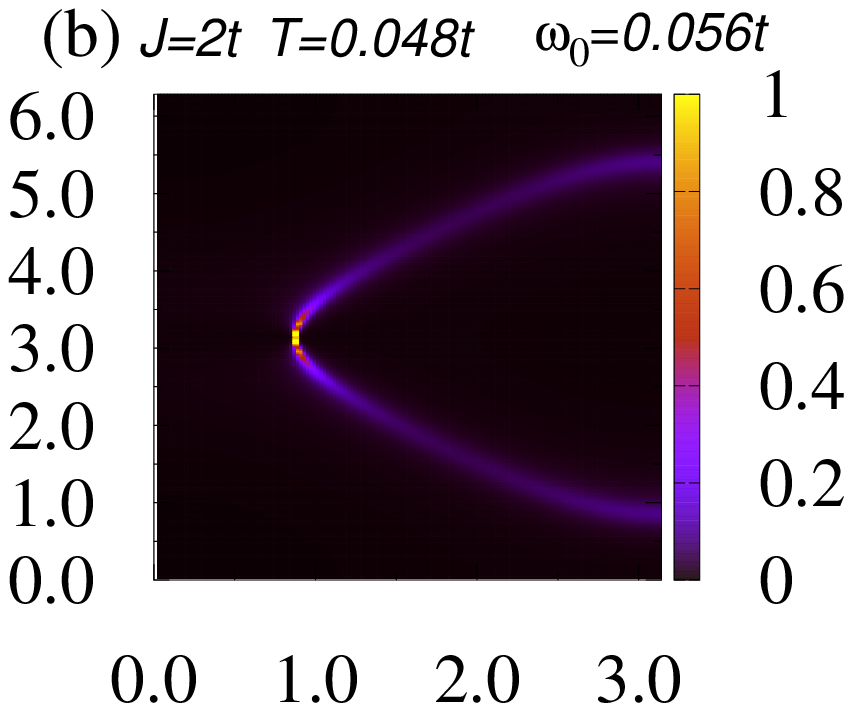}
\end{center}
\end{minipage}
\caption{(Color Online).
Momentum resolved spectral function for a given frequency $\omega_0$. 
Because the conduction electrons are effectively decoupled, we can observe the peak at $k_y=\pi$ in panel (a), signaling the emergence of the Dirac point. 
In panel (b), the effective Hamiltonian is not defective in the entire BZ although the imaginary part of the self-energy show a sharp dip around $\omega=0$ [see Fig.~\ref{fig: Sigma}(b)]
}
\label{fig: Ak_Dirac_Hybrid}
\end{figure}

We finish this section with showing that two EPs merge and change into the Dirac point. 
As $J$ decreases, the self-energy approaches to zero [see the data for $J=t$ of Fig.~\ref{fig: Sigma}]. Correspondingly, the two EPs with $\nu=1/2$ and $-1/2$ merge and change into the Dirac point observed for free-fermions[see Fig.~\ref{fig: Ak_Dirac_Hybrid}(a)]. This behavior is reasonable because these EPs are generated from the Dirac cone.
Intriguingly, however, we find that the fusion of two EPs can yield another type of defective point which has not been observed for solids.
We confirm this intriguing scenario in the low temperature region.

\subsubsection{
Fragile hybrid point and the ``Fermi loop" without defective points
}
%
Increasing the exchange interaction $J$ enhances the peak of the self-energy and shifts the position of EPs, which can lead to the fusion of two EPs at $k_x=\pi$. 
Remarkably, the fusion of two EPs generated from two distinct Dirac points yields a hybrid point, whose characterization is discussed below.
%
\begin{figure}[!h]
\begin{minipage}{0.475\hsize}
\begin{center}
\includegraphics[width=\hsize,clip]{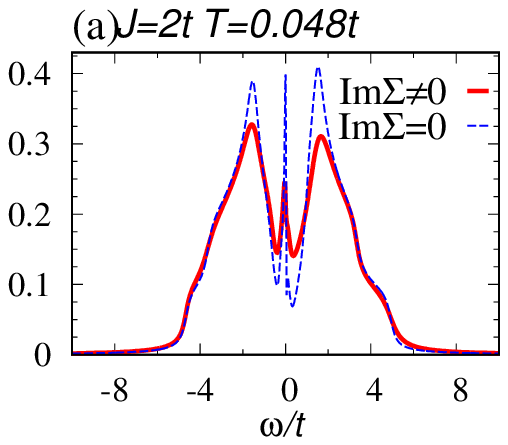}
\end{center}
\end{minipage}
\begin{minipage}{0.475\hsize}
\begin{center}
\includegraphics[width=\hsize,clip]{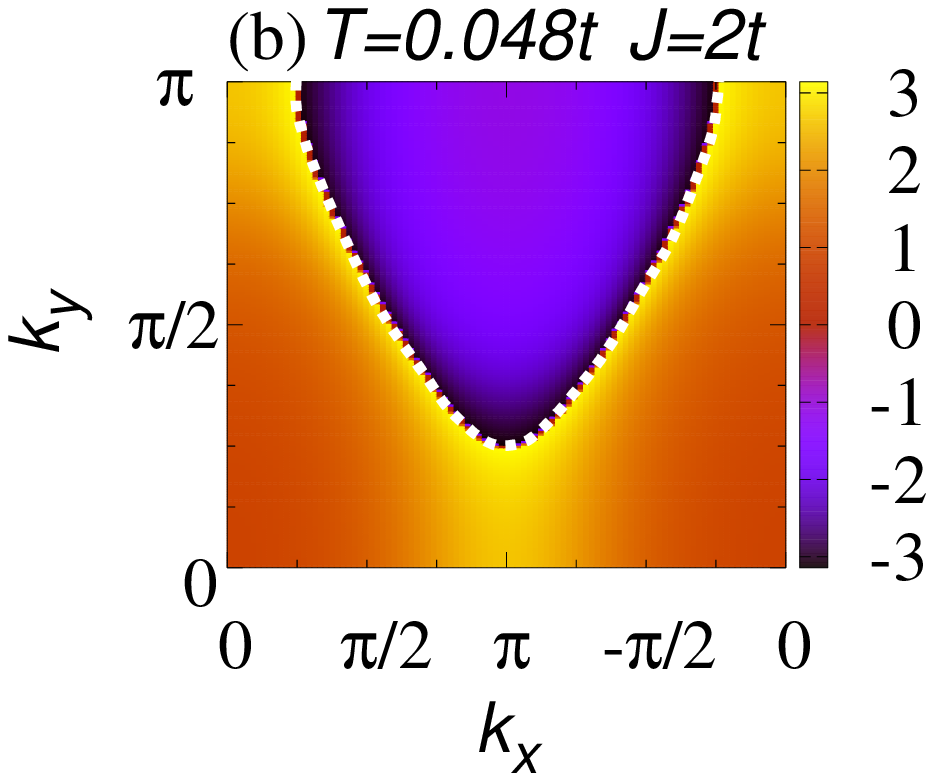}
\end{center}
\end{minipage}
\caption{
(Color Online).
Local density of states and the color map of $\Delta^2_c$ for $J=2t$ which are plotted in a similar way to Fig.~\ref{fig:LDOS_Color_J1.8}
}
\label{fig: LDOS_Color_J2}
\end{figure}
%

%
We first demonstrate the fragility of the hybrid point because of the periodicity of the Hamiltonian. 
The lattice periodicity requires $d_{x}(\bm{k})$ to be continuous and periodic in the BZ, which indicates that $d_{x}(\bm{k})^2$ has an upper bound.
Therefore, the following relation holds for any point of the BZ when the imaginary part of the self-energy takes a sufficiently large value,
%
\begin{eqnarray}
-(\mathrm{Im}\Sigma^R_{b}(\omega_0))^2 +4d^2_x(\bm{p}_0) <0,
\end{eqnarray}
and thus, the condition~(\ref{eq: EP_cond}c) is not satisfied, which results in the disappearance of the defective point with increasing temperature.
%
We note, however, that the spectrum still shows an intriguing behavior arising from the non-Hermiticity although the effective Hamiltonian is no longer defective. 
Namely, the spectral function shows the ``Fermi loop" signaling the band touching of the spectral function even in the absence of EPs  [see Fig.~\ref{fig: Ak_Dirac_Hybrid}(b)]. Because of the ``Fermi loop", the spectral weight is enhanced in the low energy region [see Fig.~\ref{fig: LDOS_Color_J2}(a)].
The origin of this ``Fermi loop" is that the line of the BZ where the generalized charge gap becomes pure imaginary forms a loop, corresponding to the absence of the defective point.

Finally, we compute the vorticity of the hybrid point. At the parameter where the two EPs merge, the condition~(\ref{eq: EP_cond}) is satisfied at $k_x=\pi$.
Corresponding to the change of Fermi arc to the Fermi loop, the branch cut of $\Delta^2_c$ also forms a loop, which can be seen even without the hybrid point [see Fig.~\ref{fig: LDOS_Color_J2}(b)]. In this case, the vorticity takes zero because any path enclosing the hybrid point crosses the branch cut twice. 

%

\subsubsection{
Antiferromagnetic instability
}

So far, we have been concerned with nonmagnetic properties. Here, we have a natural question: how does a magnetic order modify the conclusion obtained here. 

\begin{figure}[!h]
\begin{center}
\includegraphics[width=60mm,clip]{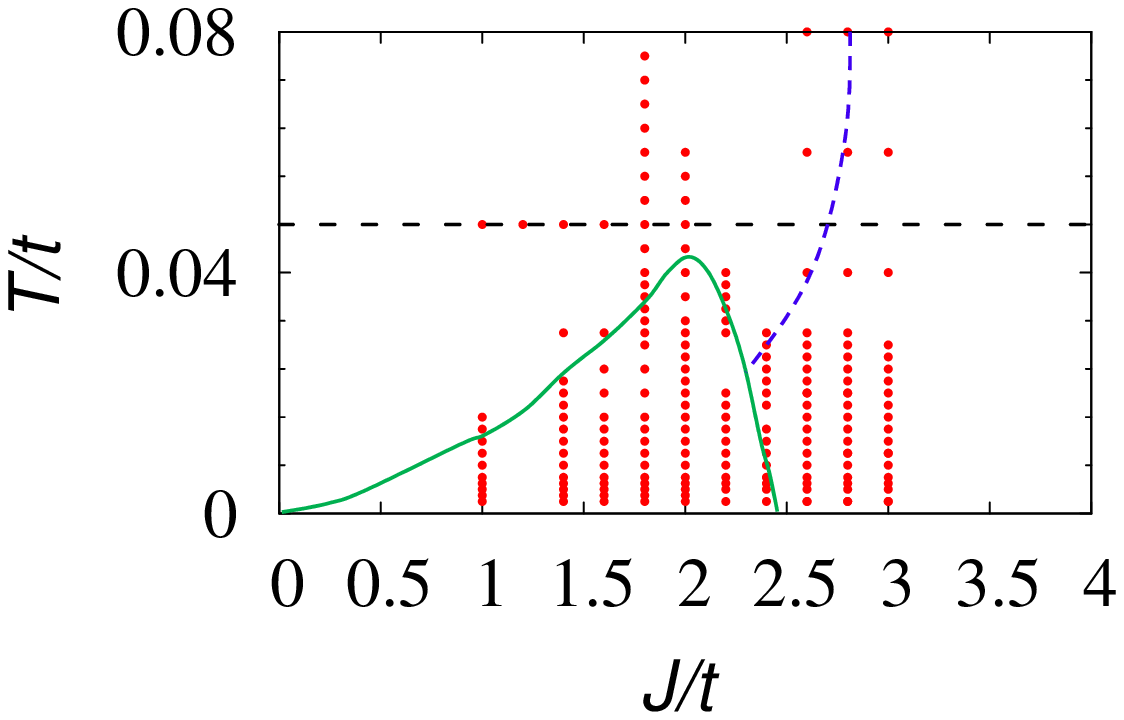}
\end{center}
\caption{(Color Online). 
Phase diagram of the spin exchange interaction $J$ vs. the temperature $T$.
The solid green line denotes the Neel temperature. The dashed blue line denotes the Kondo temperature which is estimated from saturation of the singlet-correlation between the conduction electron and the localized spin. 
The dashed horizontal line denotes the temperature $T=0.048t$.
The EPs and the ``Fermi arc" are observed above the Neel temperature.
}
\label{fig: phase}
\end{figure}

To address this question, we have performed the DMFT calculations by searching for magnetic solutions. The obtained phase diagram is shown in Fig.~\ref{fig: phase}. Red dots denote data points. It shows the typical behavior inherent to heavy fermion systems: the transition temperature of the antiferromagnetic phase increases with increasing $J$, reaches a maximum, and then decreases, entering into the Kondo insulating region. We confirm that the antiferromagnetic order changes the band structure, thereby eliminating the EPs. We note, however, that the EPs and the "Fermi loop" indeed emerge at finite temperatures higher than the Neel temperature (see Fig. 6). Thus, we conclude that the topological phenomena arising from the non-Hermitian matrix survive in our model.

\section{Summary}
%
In this paper, we have analyzed the effect of the imaginary part of the self-energy on the band structure. 
Particularly, we have analyzed the Kondo lattice model on a square lattice which shows two Dirac cones for $J=0$.

When the coupling is antiferromagnetic, increasing the exchange interaction $J$ induces the Kondo effect which results in a finite imaginary part of the self-energy in the low energy region.
Correspondingly, we have observed the emergence of EPs with vorticity $\nu=\pm 1/2$ accompanied by a ``Fermi arc" in the BZ where the charge gap $\Delta_c$ becomes pure imaginary. Along this line of the BZ, the real part of the energy is degenerate, which can be observed via the spectral weight.
Furthermore, we have observed the intriguing behavior arising from the lattice periodicity. With further increasing the spin exchange interaction $J$, the EPs are shifted and finally merge in the BZ as the Kondo effect enhances the imaginary part of the self-energy. The fusion of the two EPs, induced by the enhanced Kondo effect, yields a hybrid point which is still defective but is characterized by $\nu=0$. We have observed that this hybrid point is fragile against the slight change of the exchange interaction and vanishes with increasing $J$, leaving a ``Fermi loop" in the low energy states without any defective point.
We consider that the above intriguing phenomena, induced by correlation effects, can be observed with ARPES measurements.

\section{acknowledgements}
This work is partly supported by JSPS KAKENHI Grant No. JP15H05855, No. JP16K05501, No. JP18H01140, No. JP18H04316, and No. JP18K03511. The numerical calculations were performed on supercomputer at the ISSP in the University of Tokyo, and the SR16000 at YITP in Kyoto University.


%


\appendix

\section{Diagonalization of two-dimensional non-Hermitian matrix}
\label{sec: app_diag}
Here, we review the diagonalization of $2\times2$ non-Hermitian matrix which is, in general, given by
\begin{eqnarray}
H&=& (\bm{b}_0+i\bm{b}_1)\cdot\bm{\rho},
\end{eqnarray}
where $\rho_i$ (i=1,2,3) denotes the Pauli matrix. $\bm{b}_0,\ \bm{b}_1 \in \mathbb{R}^3$.

When the Hamiltonian is not defective, the eigenvalues are given by
\begin{eqnarray}
E_{\pm}&=&\pm \sqrt{b^2_0-b^2_1+2i\bm{b}_0\cdot\bm{b}_1}.
\end{eqnarray}

We note that the Hamiltonian is defective when the following relation holds
\begin{eqnarray}
b_0=b_1,&\quad& \bm{b}_0\cdot\bm{b}_1=0.
\end{eqnarray}

This can be seen by applying unitary transformation which maps $\bm{b}_0\cdot\bm{\rho}\to \bm{b}_0\rho_x$ and $\bm{b}_1\cdot\bm{\rho}\to b_1\rho_y$.  Under this transformation, we obtain
\begin{eqnarray}
H&=& 2b_0\left(
\begin{array}{cc}
0 & 1 \\
0 & 0
\end{array}
\right),
\end{eqnarray}
whose eigenstates cannot span the two-dimensional space.

\begin{table*}[htb]
\label{table: defective points}
  \begin{tabular}{cccc} \hline\hline
  Degenerate points & Vorticity &  Defectiveness &  creation from two EPs \\  \hline
Dirac point  & $\nu=0 $ & not defective &  EPs with $\nu=1/2$ and $-1/2$  \\
hybrid point  & $\nu=0 $ & defective &  EPs with $\nu=1/2$ and $-1/2$  \\
double exceptional point  & $\nu=1 $ & defective &  EPs with $\nu=1/2$ and $1/2$  \\
vortex point  & $\nu=1 $ & not defective &  EPs with $\nu=1/2$ and $1/2$  \\
\hline\hline
  \end{tabular}
\caption{
  Four types of degenerate points and fusion of two EPs. A Dirac point and A hybrid point can be created with two EPs with $\nu=1/2$ and $-1/2$. A double exceptional point and vortex point can be created with two EPs with $\nu=1/2$.
  }
\end{table*}
For characterization of the topological properties, the vorticity defined by Eq.~(\ref{eq: vorticity}) is used. If a defective point is characterized by vorticity of a half-integer, the defective point is denoted as an EP. 
EPs can merge and change into other types of points (see Table~\ref{table: defective points}), which is analyzed with the Hamiltonian in the continuum limit~\cite{HShen2017_non-Hermi}.
When the two EPs with $\nu=1/2$ and $-1/2$ merge, these two points changes to a Dirac point or a hybrid point. The former (latter) scenario is observed for our model in the high (low) temperature region, respectively.

\end{document}